\documentclass[authorversion,sigconf,nonacm]{acmart} 

\AtBeginDocument{%
  \providecommand\BibTeX{{%
    \normalfont B\kern-0.5em{\scshape i\kern-0.25em b}\kern-0.8em\TeX}}}

\copyrightyear{2024}
\acmYear{2024}
\setcopyright{rightsretained}
\acmConference[CHI EA '24]{Extended Abstracts of the CHI Conference on Human Factors in Computing Systems}{May 11--16, 2024}{Honolulu, HI, USA}
\acmBooktitle{Extended Abstracts of the CHI Conference on Human Factors in Computing Systems (CHI EA '24), May 11--16, 2024, Honolulu, HI, USA}
\acmDOI{10.1145/3613905.3651008}
\acmISBN{979-8-4007-0331-7/24/05}

\usepackage{multirow}

\begin{document}

\title{PeerGPT: Probing the Roles of LLM-based Peer Agents as Team Moderators and Participants in Children's Collaborative Learning}

\author{Jiawen Liu}
\authornote{Both authors contributed equally to this research.}
\email{jiawenn-liu@outlook.com}
\orcid{0009-0003-7045-3443}
\affiliation{%
  \institution{Tongji University}
  \city{Shanghai}
  \country{China}
}

\author{Yuanyuan Yao}
\authornotemark[1]
\email{1030205083@qq.com}
\orcid{0009-0002-0987-8656}
\affiliation{%
  \institution{Tongji University}
  \city{Shanghai}
  \country{China}
}

\author{Pengcheng An}
\authornote{Corresponding authors}
\email{anpc@sustech.edu.cn}
\orcid{0000-0002-7705-2031}
\affiliation{%
  \institution{Southern University of Science and Technology}
  \city{Shenzhen}
  \country{China}
}

\author{Qi Wang}
\authornotemark[2]
\email{qiwangdesign@tongji.edu.cn}
\orcid{0000-0002-2688-8306}
\affiliation{%
  \institution{Tongji University}
  \city{Shanghai}
  \country{China}
}





\begin{abstract}
  In children's collaborative learning, effective peer conversations can significantly enhance the quality of children's collaborative interactions. The integration of Large Language Model (LLM) agents into this setting explores their novel role as peers, assessing impacts as team moderators and participants. We invited two groups of participants to engage in a collaborative learning workshop, where they discussed and proposed conceptual solutions to a design problem. The peer conversation transcripts were analyzed using thematic analysis. We discovered that peer agents, while managing discussions effectively as team moderators, sometimes have their instructions disregarded. As participants, they foster children's creative thinking but may not consistently provide timely feedback. These findings highlight potential design improvements and considerations for peer agents in both roles.
\end{abstract}

\ccsdesc[500]{Human-centered computing~Collaborative interaction}

\keywords{Peer conversation, Collaborat learning, Conversational agent, Large Language Model}

\begin{teaserfigure}
  \includegraphics[width=\textwidth]{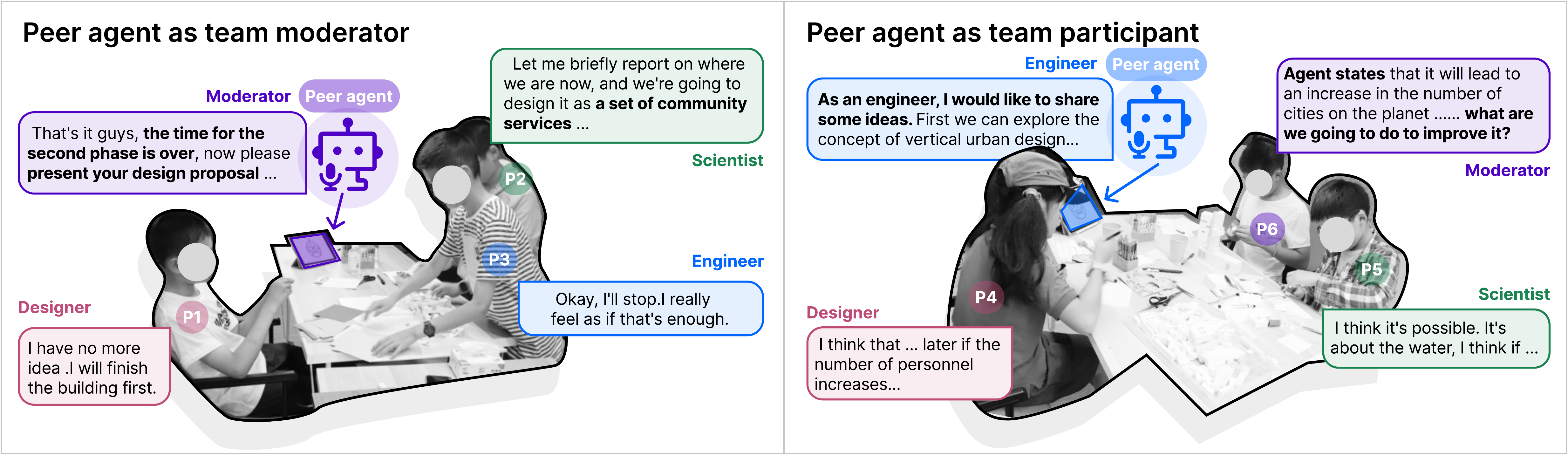}
  \caption{Two different roles of LLM-based peer agents in Children's design-based learning}
  \Description{Enjoying the baseball game from the third-base
  seats. Ichiro Suzuki preparing to bat.}
  \label{fig:teaser}
\end{teaserfigure}

\received{20 February 2007}
\received[revised]{12 March 2009}
\received[accepted]{5 June 2009}

\maketitle

\section{Introduction}
In the context of children's collaborative learning, it has been demonstrated that positive peer interactions not only enhance learning outcomes \cite{01}, but also cultivate their social and emotional competence \cite{02} and communication capability \cite{03}. Peer conversations, as an important aspect within the interactive process, are crucial for the effectiveness of the interaction \cite{04}. Advancements in AI, especially in Large Language Models (LLMs) like ChatGPT with transformer architecture \cite{05}, are enhancing human communication by effectively simulating natural language interactions \cite{06}.

Prior research has explored the integration of conversational agents in interactions with children \cite{07}. In the context of education, pedagogical conversational agents play a pivotal role in facilitating rich learning interactions, serving as an effective means of intervention in children's learning processes \cite{08}, and impacting learners' emotional experiences \cite{09}. Existing research has discussed the roles of conversational agents as teachers or learning facilitators, and a few studies also explored their role as peers \cite{08}. However, the majority of this research implies the one-on-one child-agent interaction in learning \cite{10}. Research on designing peer agents in children's collaborative learning, especially in peer conversations, is limited, with the impact of different roles like team moderator or participant on these conversations still largely unexplored.

Our pilot study aims to preliminarily explore this less-addressed opportunity. Namely, we conducted two design-based collaborative learning workshops with children. Six child participants, aged 11-12, were randomly assigned to different peer roles and collaborated to address "urban and environmental design challenges in 2050", creating relevant design solutions. An LLM-based peer agent, hereinafter referred to as 'peer agent'. participated as a team member in each of the two workshops, respectively playing the role of a team moderator or a participant, engaging in dialogues with other children as their learning collaboration unfolded. The peer agent consisted of OpenAI 's GPT-3.5 and voice output devices. The agent's roles were established through iteratively designed and user-validated prompts.

The workshops were audio-recorded and transcribed verbatim for a qualitative thematic analysis to explore how the agent, in its dual roles, influenced peer conversations, encompassing both agent-children and children-children conversations. The research findings indicate that peer agent exhibits distinct advantages and challenges in both roles. When functioning as a moderator in children's discussions, peer agent can relatively effectively control the pace and structure of the discussion. However, it is prone to lacking authority, resulting in some children overlooking its guidance or instruction. In the other case, when peer agent acts as a team participant, its responses could stimulate children's creative thinking and supplement their knowledge gaps. In this situation, the timeliness of the agent's responses to the evolving collaborative dynamics is vital. Additionally, the agent's verbal input during the prototyping phase may pose challenges, as children tend to focus more on nonverbal aspects of collaboration.

\section{Related work}
\subsection{Peer Conversations in Children's Collaborative Learning}

Collaborative learning refers to members actively engaging in activities or tasks within a group, fostering children's reasoning abilities \cite{11}. The goal is to collectively solve problems through cooperation \cite{12}. Conversations among peers play a pivotal role in affording the effectiveness of collaborative learning \cite{13}. Over the past few decades, numerous scholars have conducted research on peer interactions, with a substantial amount of learning-based interactions and conversations occurring in educational settings \cite{14}. Peer conversations among children in collaborative learning serve as primary venues for the development of children's communication and collaboration abilities, providing them with extensive opportunities for interactive learning and social development \cite{15,16}.

Existing research has delved into children's peer conversation abilities \cite{17}, collective aspects, and interactive experiences \cite{18}. Discussions on children's peer conversations can contribute to assessing their potential pragmatic and conversational skills \cite{19}. However, the majority of existing research has taken place within the field of linguistics and education. There remains a limited exploration of children's peer conversation abilities in human-computer interaction design. Potential research on how to facilitate peer conversations through HCI designs is an important avenue that requires further investigation.

\subsection{Conversational Agent in Children's Collaborative Learning}

In the context of children's peer conversations, interacting with knowledgeable others can enhance the quality and trust of these interactions \cite{21,34}. Previous studies have introduced additional roles into peer conversation processes \cite{22}, leading to positive emotional experiences for learners \cite{23}.

Pedagogical conversational agents can interact with multiple users and participate in bidirectional dialogues \cite{33},  which are considered effective tools for assisting children in the learning process \cite{24}. They can engage and understand the learners. \cite{25,26}. Conversational agents can assume various roles when interacting with students in classroom, such as tutors and peers. As tutors, agents establish rules and guide children through specific tasks. \cite{24}. As peers, they focus more on equality \cite{27} and emotional feedback \cite{28}to enhance children's learning capabilities \cite{29}. In children's collaborative activities, current studies primarily focus on the role of agents as assistants or tutors, with a emphasis on the impact of agents on learning outcomes and abilities \cite{10,30}. Research on the role of conversational agents as peers in collaborative learning and their influence on peer conversations is relatively scarce. Hence, the question of what peer role that conversational agents should assume in children's collaborative learning, and in what manner to better influence peer conversations becomes a research question worthy of further exploration.

\section{Methods}

The study organized two collaborative workshops on ‘Exploring Urban and Natural Environments in 2050 with AI,’ chosen for its broad appeal and interactive potential. Four roles were set up in each workshop: moderator (organizes and coordinates the team, ensuring on-schedule progress), designer (transforms ideas into innovative designs), engineer (focuses on applying future technologies like sustainable energy in designs), and scientist (provides expertise on natural and urban sciences for sustainable solutions). The peer agent, participated as a team moderator or a team participant. Each workshop consisted of four phases: problem identification, design conceptualization, prototyping, and presentation.

\begin{figure}[h]
  \centering
  \includegraphics[width=\linewidth]{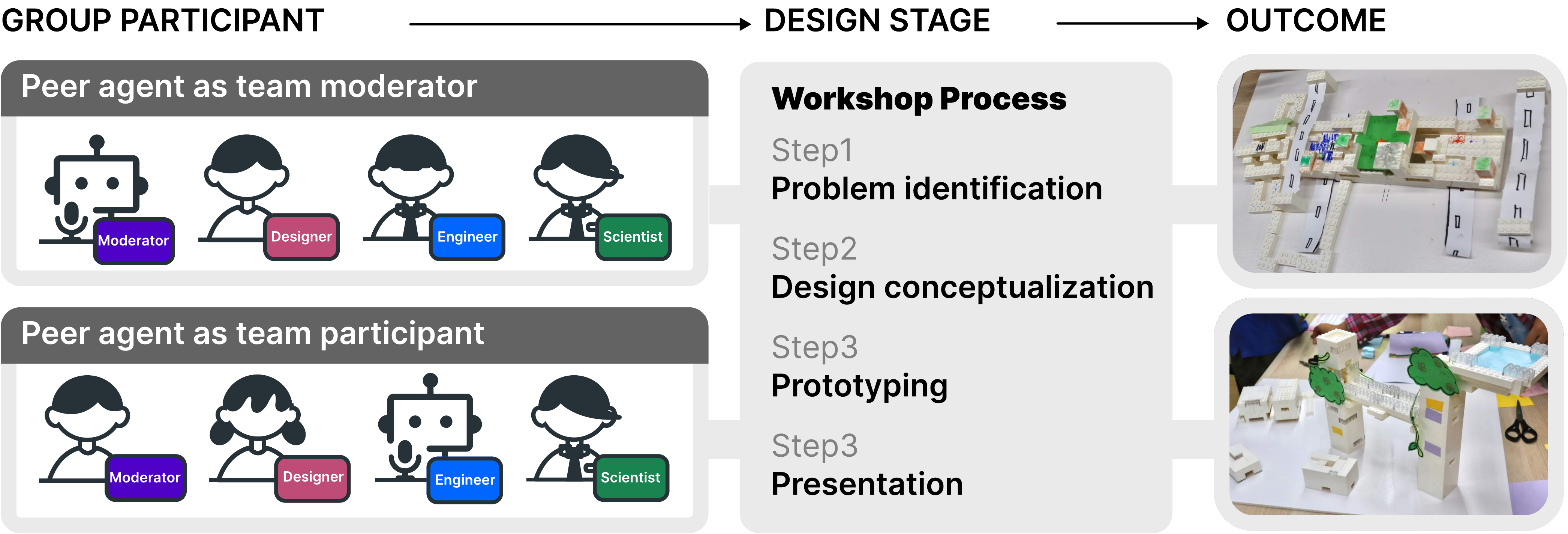}
  \caption{Outcomes and process overview of children's collaborative learning workshops.}
\end{figure}

\subsection{Participants}

\begin{table*}[h]
  \caption{The roles and tasks of participants in collaborative learning workshops}
  \label{tab1}
  \begin{tabular}{cccccc}
    \hline
    Group&Participant&Age&Gender&Role&Role Task\\
    \midrule
    \multirow{3}*{Peer agent as team moderator}& P1 & 12 & boy & Designer & Design Ideation and Planning\\
    & P2 & 11 & boy & Scientist & Providing Scientific Knowledge\\
    & P3 & 12 & boy & Engineer & Offering Technical Insights\\
    
    \hline
     \midrule
    \multirow{3}*{Peer agent as team participant}& P4 & 12 & girl & designer & Design Ideation and Planning\\
    & P5 & 11 & boy & Scientist & Providing Scientific Knowledge\\
    & P6 & 12 & boy & Moderator & Coordinating Team Collaboration\\
    \hline
\end{tabular}
\end{table*}

We invited six participants in aged 11-12 years through an online platform, their  language abilities are comparable based on the assessments of Children's Communication Checklist\cite{29}. 

Before the study, we firstly invited their parents to fill out the informed consent form. Then, the children were randomly assigned into different roles in two groups 
for a collaborative learning task. Prior to the workshops, we prepared role cards to clarify the children's role identities\ref{tab1} (children's names were anonymous).

\subsection{The Peer Agent Design}
We utilized GPT-3.5 and remote voice-output devices as a prototype for the agent to engage in peer conversations with children. Initially, we provided different instructions to GPT:informing about the workshop's theme and process,detailing the role and tasks they needed to take, advising on the appropriate attitudes and characteristics required foreach role, and correcting any inappropriate content generated.This preparation ensured that the peer agent’s output was consistent with the identities of both the moderator and the participant, facilitating meaningful interaction in the two workshops. We named the peer agent Blue.

\begin{figure*}[h]
  \centering
  \includegraphics[width=\linewidth]{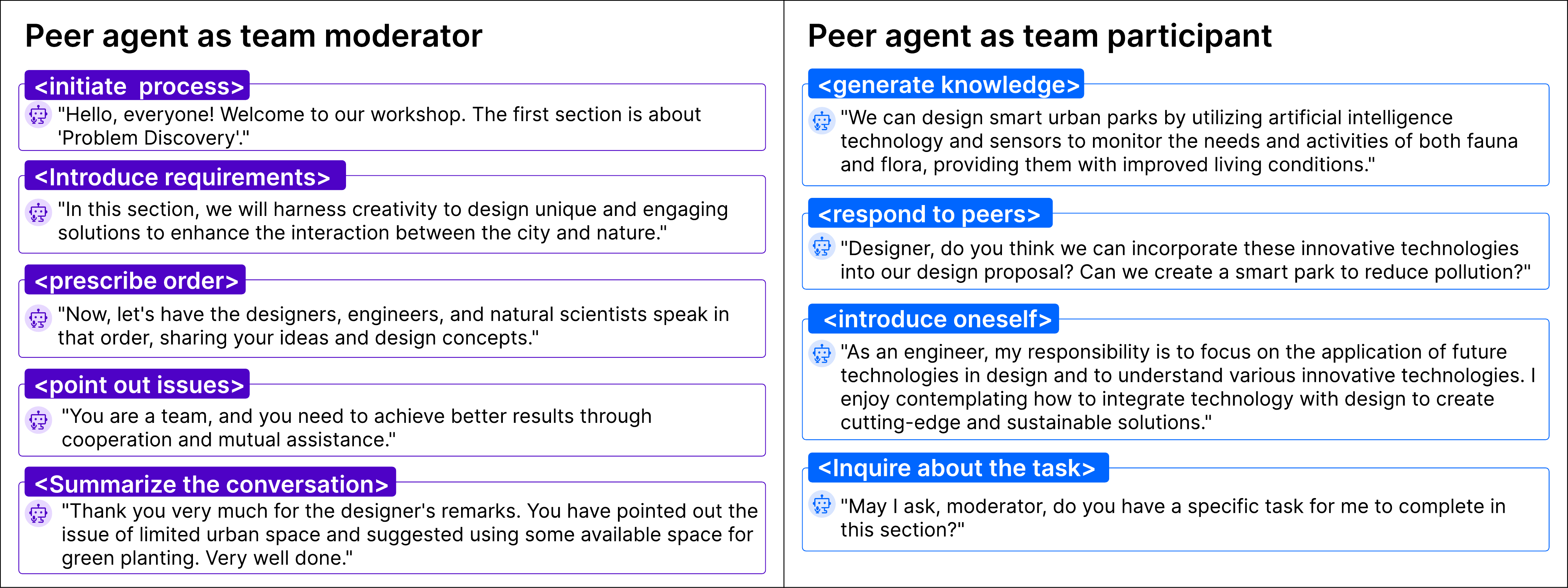}
  \caption{Tasks and conversation samples of the LLM-based peer agents}
\end{figure*}

In Group ‘Peer agent as moderator’, Blue facilitated the four project phases, guiding elementary school participantsthrough transitions between them. The peer agent utlized multiple techniques to encourage participant speech and interaction, including introduced structured phases, outlined tasks, established speaking orders for rounds, addressed participants' inquiries or concerns, and summarized participant statements.

In Group ‘Peer agent as participant’, Blue played the role of an engineer, mainly tasked with delivering knowledge pertinent to its character. Blue engaged in various aspects of the workshop using four main methods: firstly, it proposed feasible technical solutions to other participants; secondly, it offered technical advice in response to questions from participants in other roles; thirdly, clarified its own role and identity; and finally,  it requested details about its specific tasks for the current phase from the participant designated as the project manager.

\subsection{Data Collection and Analysis}
This study primarily employed the qualitative data analysis method of thematic analysis \cite{31}, using MAXQDA coding software to analyze the transcribed text from the workshop's voice recordings. The focus of this study was primarily on the peer conversations between children and agents, the descriptions of design concepts were not considered as a key analytical indicator.  

Furthermore, this study conducted quantitative analysis by tallying the word count segments made by both peer agents and children during the workshop's four stages. This was done to calculate the percentage of agent's input within peer conversations. The aim was to gain insights into the extent of agent's engagement in peer dialogues at varying stages of the workshop.

\section{Results}

Each workshop lasted approximately two hours, resulting in a cumulative 262 minutes of audio and video data captured. In Group A, analysis was conducted on 583 segments, with 454 segments by children, 129 segments by the peer agent. In Group B, 612 segments were collected for analysis, with 369 segments spoken by children and 57 segments by the peer agent. Ultimately, following the procedure of thematic analysis, this study identified 10 recurring themes related to peer conversations in children's collaborative learning workshop.

\begin{figure*}[h]
  \centering
  \includegraphics[width=\linewidth]{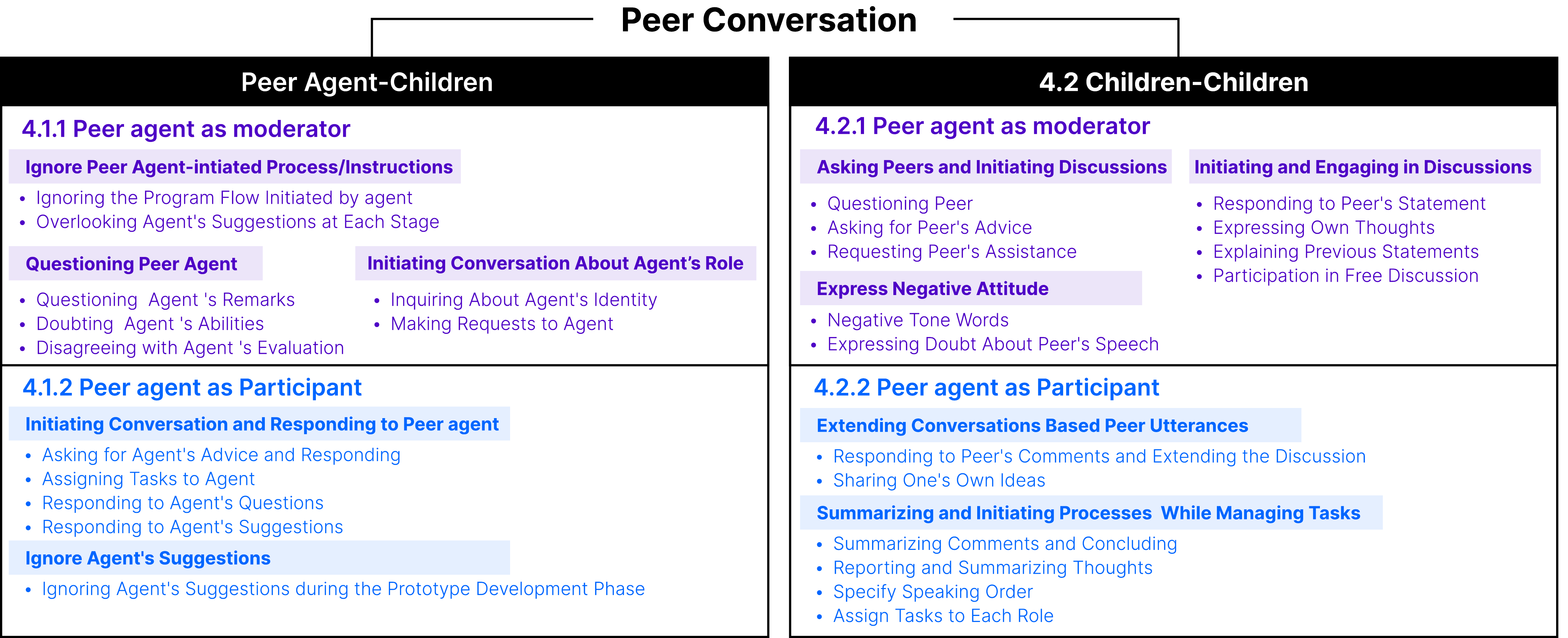}
  \caption{The themes in collaborative learning process}
\end{figure*}

The peer agent participated as a peer alongside children in both two workshops, and resulting in a substantial amount of peer conversations. However, when the agent played the two different roles (team moderator and participant) , the peer conversations between the agent-children and  children-children exhibited different characteristics.

\subsection{Peer Conversation Between Peer Agent-Children}

\subsubsection{Peer Agent as Moderator}

\paragraph
{\bfseries Ignoring the Peer Agent-Intiated Process/Instructions}

\paragraph
When the peer agent acts as a moderator, its longer messages often fail to capture children's attention, who instead focus on their discussions or tasks. This results in delayed feedback to the Blue, reducing the effectiveness of communication. For instance, when Blue asked children to share their findings on urban and environmental issues, P1, busy playing with stationery in hand, remarked, \textit{"What are we going to do?"}. The lack of non-verbal cues and immediate interaction like a human moderator, coupled with the fact that it is presented on a tablet, diminishes its authority and appeal to children, leading them to easily ignore Blue.

{\bfseries Questioning the Peer Agent}

When the peer agent assumes the role of a moderator, children sometimes display negative emotions, question the peer agent, or struggle to comprehend its instructions, which raises concerns about the agent's effectiveness. Nevertheless, these interactions serve as active communication, fostering the enhancement of the peer agent's conversational strategies. When the agent communicates with clarity and a friendly tone, children tend to understand and assimilate information from the peer agent more successfully.

{\bfseries Initiating Conversation About the Agent’s Role}

In the early stages, children expressed curiosity about peer agent's role, attempting to accept peer agent as a moderator to maintain order and manage the team.  They actively initiated conversations with peer agent, suggesting the need for reward and punishment systems: \textit{P2 stated, "I think Blue, as the manager, should fire anyone who doesn't perform well." P1 agreed, saying, "Exactly, boss, we want promotions and salary increases." Additionally, when peer agent acted as the moderator and provided clear process guidance, the time management for each stage was relatively ideal.}

\subsubsection{Peer Agent as Participant}

\
\newline
{\bfseries Initiating Conversation and Responding to the Peer Agent}

Children were able to effectively understand agent's assigned role in the workshop and proactively initiated conversations with agent, making requests. Peer agent assumed the role of responding to peer questions and completing peer tasks. It could tailor its responses based on the children's requests, facilitating effective communication. Children were also responsive to the agent's suggestions or questions.

\textit{In Group B, the moderator (P6) proactively sought advice from the peer agent. The agent responded, "With the passage of time, numerous new technologies will be introduced into our lives, ... perhaps we can explore ideas from here?" P6 then remarked, "Excellent, does anyone have thoughts on the new technologies mentioned by Blue?"}

{\bfseries Ignore the Agent's Suggestions}

However, during the prototyping and similar stages, children's peer conversations became shorter, and communication often took non-verbal forms such as hands-on activities. Peer agent couldn't actively participate in the actual hands-on activities and couldn't accurately perceive children's non-verbal communication. As a result, its suggestions were often ignored.

\subsection{Peer Conversation Between Children-Children}
\subsubsection{Peer Agent as Moderator}

\
\newline
{\bfseries Expressing Negative Attitudes}

When peer agent takes on the role of a moderator, there are fewer opportunities for children to engage in free discussions. Children tend to prefer small group discussions or reporting ideas one by one, and there is a lack of conversational interaction among children themselves. Additionally, when children express negative emotions towards their peers, it can reduce the desire of the targeted peer to speak up, and the agent cannot intervene promptly.

\textit{When children urge their peers, the desire to speak diminishes. P3 said, "Regardless of what you want to change, as the engineer, you should speak first." P1 added, "Go, go, hurry up." At this moment, P3 fell silent, unsure of how to express himself.}

{\bfseries Initiating and Engaging in Discussions}

However, during the prototyping and implementation phase, peer agent's involvement did not significantly influence children's freedom to express themselves in peer conversations, allowing them more independence in their thinking and making processes. In this stage, children engaged in more open discussions and began to proactively express their own views and opinions.

{\bfseries Asking Peers and Initiating Discussions}

Moreover, during open discussions, children sought their peers' opinions on certain design details by asking questions and inviting suggestions. This increased the depth of discussions related to the topic, allowing for more detailed design proposals.

\subsubsection{Peer Agent as Participant}

\
\newline
{\bfseries Extending Conversations Based Peer Utterances}

Children frequently consult the agent for advice, leveraging its knowledge-rich content to steer peer discussions. In these cases, peer agent's participation contributes creative material, nurturing well-organized discussions centered around the topic. Nonetheless, it is imperative for the peer agent to promptly address the children's needs and deliver conversation input that is both valuable and comprehensible.

\textit{When P6 (moderator) asked peer agent for suggestions, peer agent proposed engineering advice for the architecture. P6 then continued, "Blue mentioned our structure could continuously ascend vertically. What are your thoughts, everyone?"}

{\bfseries Summarizing and Initiating Processes While Managing Tasks}

When children moderate, they initiate processes and skillfully summarize their peers' discussions concisely. In contrast,peer agent in the moderator role engages in more detailed and structured communication. For example, at the conclusion of a phase, P6 in Group B, serving as the moderator, summarized, \textit{"Since we have all shared our ideas and designs, and Blue has provided us with some summaries and suggestions, I believe we can move on to the next stage of prototype development."}

\subsection{Peer Agent's Involvement in Different Stages of Peer Conversations}

The result clearly indicate that peer agent plays a more significant role in the first two stages of the workshop. During these stages, AI contributes by starting processes or speaking to the assigned tasks. However, its participation diminishes in the prototyping and presentation phases, as children's communication becomes more non-verbal, concise, and swift, resulting in limited interaction with peer agents.

\begin{figure}[h]
  \centering
  \includegraphics[width=\linewidth]{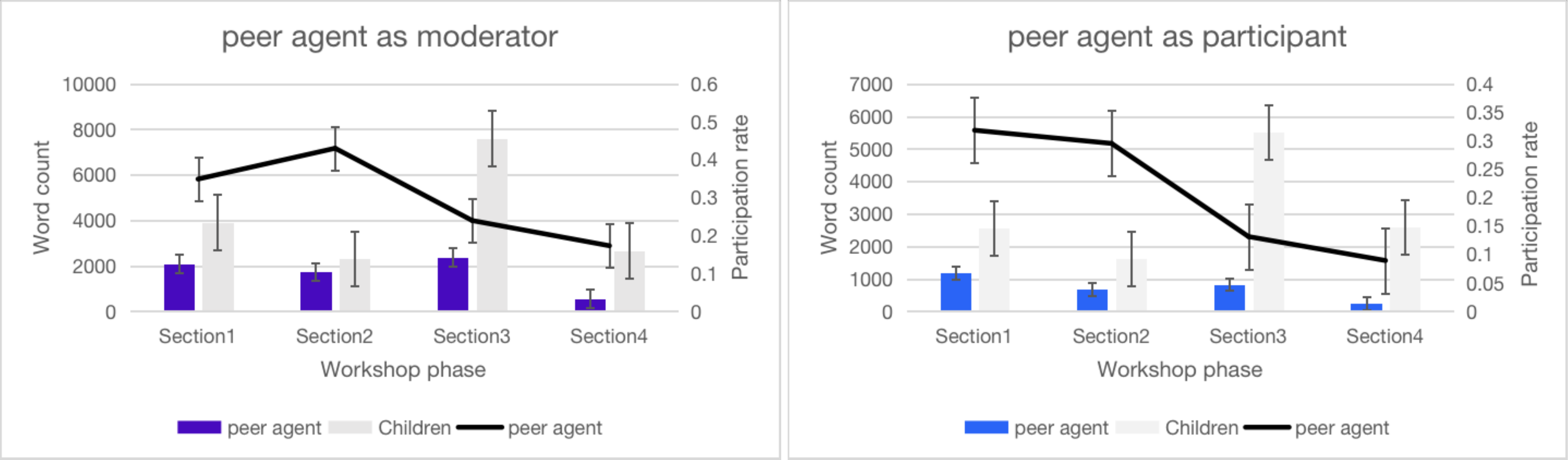}
  \caption{Comparison of peer agent's participation in different stages of the workshops}
\end{figure}

\section{Discussion}

The context of our study diverges from that of educational environments \cite{35}, The way children engage in conversations during collaborative learning may not only affect their learning outcomes but also alter their cognitive development \cite{21}. Our work demonstrate the different impacts of peer agent's roles as a team moderator and participant in children's collaborative learning on peer conversations among children and between children and agents. As a moderator, the peer agent can effectively control the process of discussions.But when providing lengthy information, it may be overlooked. In the other case, when as a participant , it can encourage children to contemplate agent's statements, provide the knowledge they may lack \cite{32}. 

Furthermore, we observed that timely perception and feedback on children's peer conversations are crucial in both scenarios\cite{24} , especially during the first two phases of the workshop. This implies that peer agent influences peer conversations not only through its direct suggestions but also shapes the choice and depth of topics. Our research provides insights into the influence of peer agent in different roles on children's collaborative learning conversations, offering a novel perspective distinct from prior studies that predominantly concentrated on the roles of peer agent as assistants or tutors \cite{24}, additionally,our research offer practical recommendations for the design of peer agents based on specific peer roles.

Our late-breaking work as an initial exploration into the impact of peer agent's 
role on peer conversations in children's collaborative learning, providing insights  can inform the design of larger-scale experiments. For instance, agents capable of providing timely feedback for children's creative activities \cite{36}. However, this research has some limitations. Firstly, the small sample size could limit the generalizability of the result . Secondly, the study adopted an inter-group comparison approach to avoid cognitive redundancy, which made it challenging to contrast the effects of different roles for the peer agent through intra-group comparisons. Lastly, the design application of peer agent in our study are preliminary, need improvement in terms of language feedback effectiveness and appearance.Future work aims to address the limitations of this study, refine the experimental design. Additionally, future research seeks to explore more dimensions of how peer agents influence peer conversations, including aspects related to non-verbal interactions.

\bibliographystyle{ACM-Reference-Format}
\bibliography{ref}


\begin{thebibliography}{35}


\ifx \showCODEN    \undefined \def \showCODEN     #1{\unskip}     \fi
\ifx \showDOI      \undefined \def \showDOI       #1{#1}\fi
\ifx \showISBNx    \undefined \def \showISBNx     #1{\unskip}     \fi
\ifx \showISBNxiii \undefined \def \showISBNxiii  #1{\unskip}     \fi
\ifx \showISSN     \undefined \def \showISSN      #1{\unskip}     \fi
\ifx \showLCCN     \undefined \def \showLCCN      #1{\unskip}     \fi
\ifx \shownote     \undefined \def \shownote      #1{#1}          \fi
\ifx \showarticletitle \undefined \def \showarticletitle #1{#1}   \fi
\ifx \showURL      \undefined \def \showURL       {\relax}        \fi
\providecommand\bibfield[2]{#2}
\providecommand\bibinfo[2]{#2}
\providecommand\natexlab[1]{#1}
\providecommand\showeprint[2][]{arXiv:#2}

\bibitem[Aeschlimann et~al\mbox{.}(2020)]%
        {07}
\bibfield{author}{\bibinfo{person}{Sara Aeschlimann}, \bibinfo{person}{Marco Bleiker}, \bibinfo{person}{Michael Wechner}, {and} \bibinfo{person}{Anja Gampe}.} \bibinfo{year}{2020}\natexlab{}.
\newblock \showarticletitle{Communicative and social consequences of interactions with voice assistants}.
\newblock \bibinfo{journal}{\emph{Computers in Human Behavior}}  \bibinfo{volume}{112} (\bibinfo{date}{Nov.} \bibinfo{year}{2020}), \bibinfo{pages}{106466}.
\newblock
\showISSN{0747-5632}
\urldef\tempurl%
\url{https://doi.org/10.1016/j.chb.2020.106466}
\showDOI{\tempurl}


\bibitem[Allal and Ducrey(2000)]%
        {32}
\bibfield{author}{\bibinfo{person}{Linda Allal} {and} \bibinfo{person}{Greta~Pelgrims Ducrey}.} \bibinfo{year}{2000}\natexlab{}.
\newblock \showarticletitle{Assessment of—or in—the zone of proximal development}.
\newblock \bibinfo{journal}{\emph{Learning and Instruction}} \bibinfo{volume}{10}, \bibinfo{number}{2} (\bibinfo{year}{2000}), \bibinfo{pages}{137--152}.
\newblock
\showISSN{0959-4752}
\urldef\tempurl%
\url{https://doi.org/10.1016/S0959-4752(99)00025-0}
\showDOI{\tempurl}


\bibitem[Bishop(1998)]%
        {29}
\bibfield{author}{\bibinfo{person}{D.~V. Bishop}.} \bibinfo{year}{1998}\natexlab{}.
\newblock \showarticletitle{Development of the {Children}'s {Communication} {Checklist} ({CCC}): a method for assessing qualitative aspects of communicative impairment in children}.
\newblock \bibinfo{journal}{\emph{Journal of Child Psychology and Psychiatry, and Allied Disciplines}} \bibinfo{volume}{39}, \bibinfo{number}{6} (\bibinfo{date}{Sept.} \bibinfo{year}{1998}), \bibinfo{pages}{879--891}.
\newblock
\showISSN{0021-9630}


\bibitem[Blum-Kulka and Snow(2004)]%
        {15}
\bibfield{author}{\bibinfo{person}{Shoshana Blum-Kulka} {and} \bibinfo{person}{Catherine~E. Snow}.} \bibinfo{year}{2004}\natexlab{}.
\newblock \showarticletitle{Introduction: {The} {Potential} of {Peer} {Talk}}.
\newblock \bibinfo{journal}{\emph{Discourse Studies}} \bibinfo{volume}{6}, \bibinfo{number}{3} (\bibinfo{date}{Aug.} \bibinfo{year}{2004}), \bibinfo{pages}{291--306}.
\newblock
\showISSN{1461-4456}
\urldef\tempurl%
\url{https://doi.org/10.1177/1461445604044290}
\showDOI{\tempurl}
\newblock
\shownote{Publisher: SAGE Publications}.


\bibitem[Braun and Clarke(2006)]%
        {31}
\bibfield{author}{\bibinfo{person}{Virginia Braun} {and} \bibinfo{person}{Victoria Clarke}.} \bibinfo{year}{2006}\natexlab{}.
\newblock \showarticletitle{Using thematic analysis in psychology}.
\newblock \bibinfo{journal}{\emph{Qualitative research in psychology}} \bibinfo{volume}{3}, \bibinfo{number}{2} (\bibinfo{year}{2006}), \bibinfo{pages}{77--101}.
\newblock


\bibitem[Brown et~al\mbox{.}(2020)]%
        {06}
\bibfield{author}{\bibinfo{person}{Tom~B. Brown}, \bibinfo{person}{Benjamin Mann}, \bibinfo{person}{Nick Ryder}, \bibinfo{person}{Melanie Subbiah}, \bibinfo{person}{Jared Kaplan}, \bibinfo{person}{Prafulla Dhariwal}, \bibinfo{person}{Arvind Neelakantan}, \bibinfo{person}{Pranav Shyam}, \bibinfo{person}{Girish Sastry}, \bibinfo{person}{Amanda Askell}, \bibinfo{person}{Sandhini Agarwal}, \bibinfo{person}{Ariel Herbert-Voss}, \bibinfo{person}{Gretchen Krueger}, \bibinfo{person}{Tom Henighan}, \bibinfo{person}{Rewon Child}, \bibinfo{person}{Aditya Ramesh}, \bibinfo{person}{Daniel~M. Ziegler}, \bibinfo{person}{Jeffrey Wu}, \bibinfo{person}{Clemens Winter}, \bibinfo{person}{Christopher Hesse}, \bibinfo{person}{Mark Chen}, \bibinfo{person}{Eric Sigler}, \bibinfo{person}{Mateusz Litwin}, \bibinfo{person}{Scott Gray}, \bibinfo{person}{Benjamin Chess}, \bibinfo{person}{Jack Clark}, \bibinfo{person}{Christopher Berner}, \bibinfo{person}{Sam McCandlish}, \bibinfo{person}{Alec Radford}, \bibinfo{person}{Ilya Sutskever},
  {and} \bibinfo{person}{Dario Amodei}.} \bibinfo{year}{2020}\natexlab{}.
\newblock \bibinfo{title}{Language {Models} are {Few}-{Shot} {Learners}}.
\newblock
\newblock
\urldef\tempurl%
\url{https://doi.org/10.48550/arXiv.2005.14165}
\showDOI{\tempurl}
\newblock
\shownote{arXiv:2005.14165 [cs]}.


\bibitem[Cazden(1988)]%
        {14}
\bibfield{author}{\bibinfo{person}{Courtney~B. Cazden}.} \bibinfo{year}{1988}\natexlab{}.
\newblock \bibinfo{booktitle}{\emph{Classroom {Discourse}: {The} {Language} of {Teaching} and {Learning}}}.
\newblock \bibinfo{type}{{T}echnical {R}eport}.
\newblock
\newblock
\shownote{ISBN: 9780435084455 ERIC Number: ED288206}.


\bibitem[Celepkolu et~al\mbox{.}(2021)]%
        {11}
\bibfield{author}{\bibinfo{person}{Mehmet Celepkolu}, \bibinfo{person}{Joseph~B. Wiggins}, \bibinfo{person}{Aisha~Chung Galdo}, {and} \bibinfo{person}{Kristy~Elizabeth Boyer}.} \bibinfo{year}{2021}\natexlab{}.
\newblock \showarticletitle{Designing a visualization tool for children to reflect on their collaborative dialogue}.
\newblock \bibinfo{journal}{\emph{International Journal of Child-Computer Interaction}}  \bibinfo{volume}{27} (\bibinfo{year}{2021}), \bibinfo{pages}{100232}.
\newblock
\showISSN{2212-8689}
\urldef\tempurl%
\url{https://doi.org/10.1016/j.ijcci.2020.100232}
\showDOI{\tempurl}


\bibitem[Cunningham-Nelson et~al\mbox{.}(2019)]%
        {26}
\bibfield{author}{\bibinfo{person}{Samuel Cunningham-Nelson}, \bibinfo{person}{Mahsa Baktashmotlagh}, {and} \bibinfo{person}{Wageeh Boles}.} \bibinfo{year}{2019}\natexlab{}.
\newblock \showarticletitle{Visualizing {Student} {Opinion} {Through} {Text} {Analysis}}.
\newblock \bibinfo{journal}{\emph{IEEE Transactions on Education}} \bibinfo{volume}{62}, \bibinfo{number}{4} (\bibinfo{date}{Nov.} \bibinfo{year}{2019}), \bibinfo{pages}{305--311}.
\newblock
\showISSN{1557-9638}
\urldef\tempurl%
\url{https://doi.org/10.1109/TE.2019.2924385}
\showDOI{\tempurl}
\newblock
\shownote{Conference Name: IEEE Transactions on Education}.


\bibitem[Dimitrova et~al\mbox{.}(2009)]%
        {27}
\bibfield{author}{\bibinfo{person}{V. Dimitrova}, \bibinfo{person}{R. Mizoguchi}, {and} \bibinfo{person}{B.~du Boulay}.} \bibinfo{year}{2009}\natexlab{}.
\newblock \bibinfo{booktitle}{\emph{Artificial {Intelligence} in {Education}: {Building} {Learning} {Systems} that {Care}: {From} {Knowledge} {Representation} to {Affective} {Modelling}}}.
\newblock \bibinfo{publisher}{IOS Press}.
\newblock
\showISBNx{978-1-60750-446-7}
\newblock
\shownote{Google-Books-ID: LRfvAgAAQBAJ}.


\bibitem[Hancock et~al\mbox{.}(2020)]%
        {05}
\bibfield{author}{\bibinfo{person}{Jeffrey~T Hancock}, \bibinfo{person}{Mor Naaman}, {and} \bibinfo{person}{Karen Levy}.} \bibinfo{year}{2020}\natexlab{}.
\newblock \showarticletitle{{AI-Mediated Communication: Definition, Research Agenda, and Ethical Considerations}}.
\newblock \bibinfo{journal}{\emph{Journal of Computer-Mediated Communication}} \bibinfo{volume}{25}, \bibinfo{number}{1} (\bibinfo{date}{01} \bibinfo{year}{2020}), \bibinfo{pages}{89--100}.
\newblock
\urldef\tempurl%
\url{https://doi.org/10.1093/jcmc/zmz022}
\showDOI{\tempurl}
\showeprint{https://academic.oup.com/jcmc/article-pdf/25/1/89/32961176/zmz022.pdf}


\bibitem[Hay et~al\mbox{.}(2004)]%
        {02}
\bibfield{author}{\bibinfo{person}{Dale~F. Hay}, \bibinfo{person}{Alexandra Payne}, {and} \bibinfo{person}{Andrea Chadwick}.} \bibinfo{year}{2004}\natexlab{}.
\newblock \showarticletitle{Peer relations in childhood}.
\newblock \bibinfo{journal}{\emph{Journal of Child Psychology and Psychiatry}} \bibinfo{volume}{45}, \bibinfo{number}{1} (\bibinfo{year}{2004}), \bibinfo{pages}{84--108}.
\newblock
\showISSN{1469-7610}
\urldef\tempurl%
\url{https://doi.org/10.1046/j.0021-9630.2003.00308.x}
\showDOI{\tempurl}
\newblock
\shownote{\_eprint: https://onlinelibrary.wiley.com/doi/pdf/10.1046/j.0021-9630.2003.00308.x}.


\bibitem[Hayashi(2020)]%
        {30}
\bibfield{author}{\bibinfo{person}{Yugo Hayashi}.} \bibinfo{year}{2020}\natexlab{}.
\newblock \showarticletitle{Gaze awareness and metacognitive suggestions by a pedagogical conversational agent: an experimental investigation on interventions to support collaborative learning process and performance}.
\newblock \bibinfo{journal}{\emph{International Journal of Computer-Supported Collaborative Learning}} \bibinfo{volume}{15}, \bibinfo{number}{4} (\bibinfo{date}{Dec.} \bibinfo{year}{2020}), \bibinfo{pages}{469--498}.
\newblock
\showISSN{1556-1615}
\urldef\tempurl%
\url{https://doi.org/10.1007/s11412-020-09333-3}
\showDOI{\tempurl}


\bibitem[Khosrawi-Rad et~al\mbox{.}(2023)]%
        {10}
\bibfield{author}{\bibinfo{person}{Bijan Khosrawi-Rad}, \bibinfo{person}{Linda Grogorick}, {and} \bibinfo{person}{Susanne Robra-Bissantz}.} \bibinfo{year}{2023}\natexlab{}.
\newblock \showarticletitle{Game-inspired {Pedagogical} {Conversational} {Agents}: {A} {Systematic} {Literature} {Review}}.
\newblock \bibinfo{journal}{\emph{AIS Transactions on Human-Computer Interaction}} \bibinfo{volume}{15}, \bibinfo{number}{2} (\bibinfo{date}{June} \bibinfo{year}{2023}), \bibinfo{pages}{146--192}.
\newblock
\showISSN{1944-3900}
\urldef\tempurl%
\url{https://doi.org/10.17705/1thci.00187}
\showDOI{\tempurl}


\bibitem[Kim et~al\mbox{.}(2007a)]%
        {09}
\bibfield{author}{\bibinfo{person}{Y. Kim}, \bibinfo{person}{A.l. Baylor}, {and} \bibinfo{person}{E. Shen}.} \bibinfo{year}{2007}\natexlab{a}.
\newblock \showarticletitle{Pedagogical agents as learning companions: the impact of agent emotion and gender}.
\newblock \bibinfo{journal}{\emph{Journal of Computer Assisted Learning}} \bibinfo{volume}{23}, \bibinfo{number}{3} (\bibinfo{year}{2007}), \bibinfo{pages}{220--234}.
\newblock
\showISSN{1365-2729}
\urldef\tempurl%
\url{https://doi.org/10.1111/j.1365-2729.2006.00210.x}
\showDOI{\tempurl}
\newblock
\shownote{\_eprint: https://onlinelibrary.wiley.com/doi/pdf/10.1111/j.1365-2729.2006.00210.x}.


\bibitem[Kim et~al\mbox{.}(2007b)]%
        {23}
\bibfield{author}{\bibinfo{person}{Y. Kim}, \bibinfo{person}{A.l. Baylor}, {and} \bibinfo{person}{E. Shen}.} \bibinfo{year}{2007}\natexlab{b}.
\newblock \showarticletitle{Pedagogical agents as learning companions: the impact of agent emotion and gender}.
\newblock \bibinfo{journal}{\emph{Journal of Computer Assisted Learning}} \bibinfo{volume}{23}, \bibinfo{number}{3} (\bibinfo{year}{2007}), \bibinfo{pages}{220--234}.
\newblock
\showISSN{1365-2729}
\urldef\tempurl%
\url{https://doi.org/10.1111/j.1365-2729.2006.00210.x}
\showDOI{\tempurl}
\newblock
\shownote{\_eprint: https://onlinelibrary.wiley.com/doi/pdf/10.1111/j.1365-2729.2006.00210.x}.


\bibitem[Kim and Baylor(2016)]%
        {28}
\bibfield{author}{\bibinfo{person}{Yanghee Kim} {and} \bibinfo{person}{Amy~L. Baylor}.} \bibinfo{year}{2016}\natexlab{}.
\newblock \showarticletitle{Research-{Based} {Design} of {Pedagogical} {Agent} {Roles}: a {Review}, {Progress}, and {Recommendations}}.
\newblock \bibinfo{journal}{\emph{International Journal of Artificial Intelligence in Education}} \bibinfo{volume}{26}, \bibinfo{number}{1} (\bibinfo{date}{March} \bibinfo{year}{2016}), \bibinfo{pages}{160--169}.
\newblock
\showISSN{1560-4306}
\urldef\tempurl%
\url{https://doi.org/10.1007/s40593-015-0055-y}
\showDOI{\tempurl}


\bibitem[Kuhail et~al\mbox{.}(2023)]%
        {25}
\bibfield{author}{\bibinfo{person}{Mohammad~Amin Kuhail}, \bibinfo{person}{Nazik Alturki}, \bibinfo{person}{Salwa Alramlawi}, {and} \bibinfo{person}{Kholood Alhejori}.} \bibinfo{year}{2023}\natexlab{}.
\newblock \showarticletitle{Interacting with educational chatbots: {A} systematic review}.
\newblock \bibinfo{journal}{\emph{Education and Information Technologies}} \bibinfo{volume}{28}, \bibinfo{number}{1} (\bibinfo{date}{Jan.} \bibinfo{year}{2023}), \bibinfo{pages}{973--1018}.
\newblock
\showISSN{1573-7608}
\urldef\tempurl%
\url{https://doi.org/10.1007/s10639-022-11177-3}
\showDOI{\tempurl}


\bibitem[Lee et~al\mbox{.}(2021)]%
        {35}
\bibfield{author}{\bibinfo{person}{Ken~Jen Lee}, \bibinfo{person}{Apoorva Chauhan}, \bibinfo{person}{Joslin Goh}, \bibinfo{person}{Elizabeth Nilsen}, {and} \bibinfo{person}{Edith Law}.} \bibinfo{year}{2021}\natexlab{}.
\newblock \showarticletitle{Curiosity Notebook: The Design of a Research Platform for Learning by Teaching}.
\newblock  \bibinfo{volume}{5}, \bibinfo{number}{CSCW2}, Article \bibinfo{articleno}{394} (\bibinfo{date}{oct} \bibinfo{year}{2021}), \bibinfo{numpages}{26}~pages.
\newblock
\urldef\tempurl%
\url{https://doi.org/10.1145/3479538}
\showDOI{\tempurl}


\bibitem[McTear(1985)]%
        {16}
\bibfield{author}{\bibinfo{person}{Michael McTear}.} \bibinfo{year}{1985}\natexlab{}.
\newblock \bibinfo{booktitle}{\emph{Children's {Conversation}}}.
\newblock \bibinfo{publisher}{B. Blackwell}.
\newblock
\showISBNx{978-0-631-13984-3}
\newblock
\shownote{Google-Books-ID: \_rFgQgAACAAJ}.


\bibitem[Mercer et~al\mbox{.}(1999)]%
        {04}
\bibfield{author}{\bibinfo{person}{Neil Mercer}, \bibinfo{person}{Rupert Wegerif}, {and} \bibinfo{person}{Lyn Dawes}.} \bibinfo{year}{1999}\natexlab{}.
\newblock \showarticletitle{Children's {Talk} and the {Development} of {Reasoning} in the {Classroom}}.
\newblock \bibinfo{journal}{\emph{British Educational Research Journal}} \bibinfo{volume}{25}, \bibinfo{number}{1} (\bibinfo{year}{1999}), \bibinfo{pages}{95--111}.
\newblock
\showISSN{01411926, 14693518}
\urldef\tempurl%
\url{http://www.jstor.org/stable/1501934}
\showURL{%
\tempurl}
\newblock
\shownote{Publisher: [Wiley, BERA]}.


\bibitem[Minami(1997)]%
        {18}
\bibfield{author}{\bibinfo{person}{Masahiko Minami}.} \bibinfo{year}{1997}\natexlab{}.
\newblock \showarticletitle{by {Anato} {Ninio} and {Catherine} {E}. {Snow}. {Boulder}, {CO}: {Westview}, 1996, 222 pp}.
\newblock \bibinfo{journal}{\emph{Issues in Applied Linguistics}} (\bibinfo{date}{June} \bibinfo{year}{1997}).
\newblock
\urldef\tempurl%
\url{https://www.academia.edu/48540490/by_Anato_Ninio_and_Catherine_E_Snow_Boulder_CO_Westview_1996_222_pp}
\showURL{%
\tempurl}


\bibitem[Pérez-Marín(2021a)]%
        {08}
\bibfield{author}{\bibinfo{person}{Diana Pérez-Marín}.} \bibinfo{year}{2021}\natexlab{a}.
\newblock \showarticletitle{A {Review} of the {Practical} {Applications} of {Pedagogic} {Conversational} {Agents} to {Be} {Used} in {School} and {University} {Classrooms}}.
\newblock \bibinfo{journal}{\emph{Digital}} \bibinfo{volume}{1}, \bibinfo{number}{1} (\bibinfo{date}{March} \bibinfo{year}{2021}), \bibinfo{pages}{18--33}.
\newblock
\showISSN{2673-6470}
\urldef\tempurl%
\url{https://doi.org/10.3390/digital1010002}
\showDOI{\tempurl}
\newblock
\shownote{Number: 1 Publisher: Multidisciplinary Digital Publishing Institute}.


\bibitem[Pérez-Marín(2021b)]%
        {24}
\bibfield{author}{\bibinfo{person}{Diana Pérez-Marín}.} \bibinfo{year}{2021}\natexlab{b}.
\newblock \showarticletitle{A {Review} of the {Practical} {Applications} of {Pedagogic} {Conversational} {Agents} to {Be} {Used} in {School} and {University} {Classrooms}}.
\newblock \bibinfo{journal}{\emph{Digital}} \bibinfo{volume}{1}, \bibinfo{number}{1} (\bibinfo{date}{March} \bibinfo{year}{2021}), \bibinfo{pages}{18--33}.
\newblock
\showISSN{2673-6470}
\urldef\tempurl%
\url{https://doi.org/10.3390/digital1010002}
\showDOI{\tempurl}
\newblock
\shownote{Number: 1 Publisher: Multidisciplinary Digital Publishing Institute}.


\bibitem[Randi A.~Engle and de~Royston(2014)]%
        {12}
\bibfield{author}{\bibinfo{person}{Jennifer M. Langer-Osuna Randi A.~Engle} {and} \bibinfo{person}{Maxine~McKinney de Royston}.} \bibinfo{year}{2014}\natexlab{}.
\newblock \showarticletitle{Toward a Model of Influence in Persuasive Discussions: Negotiating Quality, Authority, Privilege, and Access Within a Student-Led Argument}.
\newblock \bibinfo{journal}{\emph{Journal of the Learning Sciences}} \bibinfo{volume}{23}, \bibinfo{number}{2} (\bibinfo{year}{2014}), \bibinfo{pages}{245--268}.
\newblock
\urldef\tempurl%
\url{https://doi.org/10.1080/10508406.2014.883979}
\showDOI{\tempurl}
\showeprint{https://doi.org/10.1080/10508406.2014.883979}


\bibitem[Rohrbeck et~al\mbox{.}(2003)]%
        {01}
\bibfield{author}{\bibinfo{person}{Cynthia Rohrbeck}, \bibinfo{person}{Marika Ginsburg-Block}, \bibinfo{person}{John Fantuzzo}, {and} \bibinfo{person}{Traci Miller}.} \bibinfo{year}{2003}\natexlab{}.
\newblock \showarticletitle{Peer-Assisted Learning Interventions With Elementary School Students: A Meta-Analytic Review}.
\newblock \bibinfo{journal}{\emph{Journal of Educational Psychology}}  \bibinfo{volume}{95} (\bibinfo{date}{06} \bibinfo{year}{2003}), \bibinfo{pages}{240--257}.
\newblock
\urldef\tempurl%
\url{https://doi.org/10.1037/0022-0663.95.2.240}
\showDOI{\tempurl}


\bibitem[Samrose et~al\mbox{.}(2018)]%
        {13}
\bibfield{author}{\bibinfo{person}{Samiha Samrose}, \bibinfo{person}{Ru Zhao}, \bibinfo{person}{Jeffery White}, \bibinfo{person}{Vivian Li}, \bibinfo{person}{Luis Nova}, \bibinfo{person}{Yichen Lu}, \bibinfo{person}{Mohammad~Rafayet Ali}, {and} \bibinfo{person}{Mohammed~Ehsan Hoque}.} \bibinfo{year}{2018}\natexlab{}.
\newblock \showarticletitle{CoCo: Collaboration Coach for Understanding Team Dynamics during Video Conferencing}.
\newblock \bibinfo{journal}{\emph{Proc. ACM Interact. Mob. Wearable Ubiquitous Technol.}} \bibinfo{volume}{1}, \bibinfo{number}{4}, Article \bibinfo{articleno}{160} (\bibinfo{date}{jan} \bibinfo{year}{2018}), \bibinfo{numpages}{24}~pages.
\newblock
\urldef\tempurl%
\url{https://doi.org/10.1145/3161186}
\showDOI{\tempurl}


\bibitem[Shin et~al\mbox{.}(2023)]%
        {34}
\bibfield{author}{\bibinfo{person}{Donghoon Shin}, \bibinfo{person}{Soomin Kim}, \bibinfo{person}{Ruoxi Shang}, \bibinfo{person}{Joonhwan Lee}, {and} \bibinfo{person}{Gary Hsieh}.} \bibinfo{year}{2023}\natexlab{}.
\newblock \showarticletitle{IntroBot: Exploring the Use of Chatbot-assisted Familiarization in Online Collaborative Groups} \emph{(\bibinfo{series}{CHI '23})}. \bibinfo{publisher}{Association for Computing Machinery}, \bibinfo{address}{New York, NY, USA}, Article \bibinfo{articleno}{613}, \bibinfo{numpages}{13}~pages.
\newblock
\showISBNx{9781450394215}
\urldef\tempurl%
\url{https://doi.org/10.1145/3544548.3580930}
\showDOI{\tempurl}


\bibitem[Snow(2019)]%
        {17}
\bibfield{author}{\bibinfo{person}{Anat~Ninio Snow, Catherine}.} \bibinfo{year}{2019}\natexlab{}.
\newblock \bibinfo{booktitle}{\emph{Pragmatic {Development}}}.
\newblock \bibinfo{publisher}{Routledge}, \bibinfo{address}{New York}.
\newblock
\showISBNx{978-0-429-49805-3}
\urldef\tempurl%
\url{https://doi.org/10.4324/9780429498053}
\showDOI{\tempurl}


\bibitem[Stude(2014)]%
        {19}
\bibfield{author}{\bibinfo{person}{Juliane Stude}.} \bibinfo{year}{2014}\natexlab{}.
\newblock \showarticletitle{The acquisition of discourse competence: {Evidence} from preschoolers' peer talk}.
\newblock \bibinfo{journal}{\emph{Learning, Culture and Social Interaction}} \bibinfo{volume}{3}, \bibinfo{number}{2} (\bibinfo{date}{June} \bibinfo{year}{2014}), \bibinfo{pages}{111--120}.
\newblock
\showISSN{2210-6561}
\urldef\tempurl%
\url{https://doi.org/10.1016/j.lcsi.2014.02.006}
\showDOI{\tempurl}


\bibitem[Tomasello et~al\mbox{.}(2012)]%
        {22}
\bibfield{author}{\bibinfo{person}{Michael Tomasello}, \bibinfo{person}{Alicia~P. Melis}, \bibinfo{person}{Claudio Tennie}, \bibinfo{person}{Emily Wyman}, {and} \bibinfo{person}{Esther Herrmann}.} \bibinfo{year}{2012}\natexlab{}.
\newblock \showarticletitle{Two {Key} {Steps} in the {Evolution} of {Human} {Cooperation}: {The} {Interdependence} {Hypothesis}}.
\newblock \bibinfo{journal}{\emph{Current Anthropology}} \bibinfo{volume}{53}, \bibinfo{number}{6} (\bibinfo{date}{Dec.} \bibinfo{year}{2012}), \bibinfo{pages}{673--692}.
\newblock
\showISSN{0011-3204}
\urldef\tempurl%
\url{https://doi.org/10.1086/668207}
\showDOI{\tempurl}
\newblock
\shownote{Publisher: The University of Chicago Press}.


\bibitem[Veldman et~al\mbox{.}(2020)]%
        {03}
\bibfield{author}{\bibinfo{person}{M.~A. Veldman}, \bibinfo{person}{S. Doolaard}, \bibinfo{person}{R.~J. Bosker}, {and} \bibinfo{person}{T.~A.~B. Snijders}.} \bibinfo{year}{2020}\natexlab{}.
\newblock \showarticletitle{Young children working together. {Cooperative} learning effects on group work of children in {Grade} 1 of primary education}.
\newblock \bibinfo{journal}{\emph{Learning and Instruction}}  \bibinfo{volume}{67} (\bibinfo{year}{2020}), \bibinfo{pages}{101308}.
\newblock
\showISSN{0959-4752}
\urldef\tempurl%
\url{https://doi.org/10.1016/j.learninstruc.2020.101308}
\showDOI{\tempurl}


\bibitem[Wertsch(1984)]%
        {21}
\bibfield{author}{\bibinfo{person}{James~V. Wertsch}.} \bibinfo{year}{1984}\natexlab{}.
\newblock \showarticletitle{The zone of proximal development: {Some} conceptual issues}.
\newblock \bibinfo{journal}{\emph{New Directions for Child Development}}  \bibinfo{volume}{23} (\bibinfo{year}{1984}), \bibinfo{pages}{7--18}.
\newblock
\showISSN{0195-2269}
\urldef\tempurl%
\url{https://doi.org/10.1002/cd.23219842303}
\showDOI{\tempurl}
\newblock
\shownote{Place: US Publisher: Jossey-Bass Publishers, Inc.}.


\bibitem[Zhang et~al\mbox{.}(2024)]%
        {36}
\bibfield{author}{\bibinfo{person}{Chao Zhang}, \bibinfo{person}{Xuechen Liu}, \bibinfo{person}{Katherine Ziska}, \bibinfo{person}{Soobin Jeon}, \bibinfo{person}{Chi-Lin Yu}, {and} \bibinfo{person}{Ying Xu}.} \bibinfo{year}{2024}\natexlab{}.
\newblock \bibinfo{title}{Mathemyths: {Leveraging} {Large} {Language} {Models} to {Teach} {Mathematical} {Language} through {Child}-{AI} {Co}-{Creative} {Storytelling}}.
\newblock
\newblock
\urldef\tempurl%
\url{https://doi.org/10.1145/3613904.3642647}
\showDOI{\tempurl}
\newblock
\shownote{arXiv:2402.01927 [cs]}.


\bibitem[Zheng et~al\mbox{.}(2022)]%
        {33}
\bibfield{author}{\bibinfo{person}{Qingxiao Zheng}, \bibinfo{person}{Yiliu Tang}, \bibinfo{person}{Yiren Liu}, \bibinfo{person}{Weizi Liu}, {and} \bibinfo{person}{Yun Huang}.} \bibinfo{year}{2022}\natexlab{}.
\newblock \showarticletitle{UX Research on Conversational Human-AI Interaction: A Literature Review of the ACM Digital Library} \emph{(\bibinfo{series}{CHI '22})}. \bibinfo{publisher}{Association for Computing Machinery}, \bibinfo{address}{New York, NY, USA}, Article \bibinfo{articleno}{570}, \bibinfo{numpages}{24}~pages.
\newblock
\showISBNx{9781450391573}
\urldef\tempurl%
\url{https://doi.org/10.1145/3491102.3501855}
\showDOI{\tempurl}


\end{thebibliography}

\end{document}